# PREPRINTS

## Graphene/SiO$_2$ nanocomposites: the enhancement of photocatalytic and biomedical activity of SiO$_2$ nanoparticles by graphene

**Aqsa Arshad[1,2,a], Javed Iqbal[3,a], Qaisar Mansoor[4], Ishaq Ahmed[5]**

[1]*Department of Physics, International Islamic University, Islamabad, Pakistan.*

[2]*Department of Physics, Durham University, South Road Durham DH1 3LE, United Kingdom.*

[3] *Laboratory of Nanoscience and Technology, Department of Physics, Quaid i Azam University, Islamabad, Pakistan.*

[4]*Institute of Biomedical and Genetic Engineering (IBGE), Islamabad, Pakistan.*

[5]*Experimental Physics Labs, Professor Abdus Salam Centre of Physics, Islamabad, Pakistan.*

[a] Corresponding author(s): aqsa.arshad@iiu.edu.pk ; javed.saggu@qau.edu.pk

**ABSTRACT:** The exceptional conducting nature of graphene makes it a viable candidate for enhancing the effectiveness of photocatalytic and biomedical nanomaterials. Herein, the immobilization of monodispersed silicon dioxide ($SiO_2$) nanoparticles on multiple graphene layers is demonstrated for intercalation of graphene nanoplatelets (GNPs). Interestingly, the loading of graphene nanoplatelets with $SiO_2$ nanoparticles enhances the photocatalytic efficiency from 46% to 99%. For biomedical applications, it is found that 75% of Gram positive and 50% of Gram negative bacteria have been killed, hence bacterial proliferation is significantly restricted. Further, the cytotoxicity study reveals that the synthesised nanocomposites are non-toxic for both normal (HCEC) and cancerous (MCF-7, HEp-2) cell lines which signifies their potential as carriers for drug delivery. The prepared nanocomposites with controlled amount of carbon in the form of graphene can be employed for photocatalysis based waste water remediation, biomedicine and nano drug delivery.

KEYWORDS: *Toxic nanomaterials; Graphene nanocomposites; Photocatalysis; Waste water treatment; Cytotoxicity; Biomedicine*

## I.   INTRODUCTION

Health and ecological issues arising due to microorganism and rapid industrialization have posed the excessive risk of diseases and water contamination. These problems have enforced researchers to develop new environment friendly materials to cope with worldwide health hazards and treatment of waste water. Nowadays, finding solutions to clean the water contaminated by dyes is prime focus of research. At the same time, synthesis of bactericidal materials[1] is equally significant for cleaning infected water, food packaging, hospital implants, and dentistry instruments.[2-5] Furthermore, in the health sector, cancer is the most threatening and incurable disease known till now.[6] Therefore, the chemotherapeutic treatments are essential and such treatments require nontoxic nanosized carriers for efficient drug delivery.[7]

Metal and non-metal oxides have been extensively studied for photocatalysis and biomedical applications.[8-12] Among them, $SiO_2$ has the prime importance as it is the promising candidate for future development of catalysts, adsorbents, nanodrug carriers and biomolecular transport agents.[13-20] However, some demerits of $SiO_2$ restrict its efficiency in photocatalysis and biomedicine, which are: the wide bandgap (~5eV), aggregation of NPs, quick charge carrier's recombination and low surface area. These factors contribute towards its inert behaviour towards many catalytic processes and it shows only a slight catalytic activity under UV irradiation.[21,22] To improve its photocatalytic performance, it is highly desirable to combine carbonaceous materials with $SiO_2$, as various studies indeed describe their efficacy for remediation of contaminated water.[23-26] The latest addition to carbonaceous materials is graphene which is a perfectly two dimensional, $sp^2$ hybridized carbon. The ultrafast Dirac Fermions (near the k and k' points) render exceptional transport properties in graphene.[27] The enticing features that make graphene ideal for various applications are its high electrical and thermal conductivities, large specific surface area, good chemical stability and outstanding mechanical properties.[28-30] These features pave the way for incorporation of graphene in hybrid materials (constituted by its combination with metals, ceramics, polymers, and chalcogenides).[31-36] Harnessing the good electronic transport and other physical properties of graphene with those of ceramics can significantly enhance their performance in photocatalysis and biomedical applications.[37-39] For instance, the activity of a nanomaterial to remove dye and bacterial contamination is mainly influenced by surface area, surface roughness and functionalization.[40] $SiO_2$ NPs decorated graphene nanoplatelets (GNPs) have hybrid properties of both nanomaterials e.g., improved interfacial contact (leading to large surface area), development of conducting pathways, and suppression of charge recombination. Incorporating these features via formation of graphene/SiO$_2$ nanocomposites provides a superior channel for enhancement of the photocatalytic performance as compared to other carbonaceous nanocomposites or SiO$_2$ alone.[41] The combined assembly of SiO$_2$ and graphene also makes possible the covalent and non-covalent attachments of drugs on graphene by $\pi$–$\pi$ interactions. This makes it a suitable nano-carrier for delivering drug to target cancer sites.[42] This outcome can't be achieved by SiO$_2$ NPs alone. Thus, these merits make graphene/SiO$_2$ nanocomposites, potentially new candidates for averting the ever-growing health and environmental risks associated with contaminated water. Graphene family based nanocomposites with SiO$_2$ have recently been studied in detail for various applications e.g., liquid chromatography, fluorescence enhancement, super hydrophilic coatings.[43-47] Despite these few reports, there is still a lot of room to probe various aspects of graphene/SiO$_2$ nanocomposites. For example, health and environmental impacts of graphene/SiO$_2$ nanocomposite need to be thoroughly evaluated before employing it for potential applications. In this work, photocatalytic activity under UV light illumination for a model organic dye methyl orange (MO), antibacterial features of graphene/SiO$_2$ nanocomposite for three model bacteria and the cytotoxicity of graphene/SiO$_2$ nanocomposite for normal (HCEC) and two cancerous cell lines (MCF-7, HEp-2) have been evaluated.

For the first time, this study explores the impact of graphene loading on the role of SiO$_2$ in the fast photodegradation of MO and inhibition of the bacterial growth. Moreover, the cytotoxicity evaluations suggest the possible use of graphene/SiO$_2$ nanocomposites as a nano carrier for drug delivery and bio-imaging in cancer treatment procedures.

## II.   EXPERIMENTAL METHODS

### A.  SYNTHESIS OF GRAPHENE/SiO$_2$ NANOCOMPOSITES

Tetraethyl ortho silicate (TEOS) (99%, Fluka), liquid ammonia (32%), ethyl alcohol (99%, Merck), GNPs (100%, KNano) and distilled water were used in the fabrication of monodispersed $SiO_2$ particles and graphene/$SiO_2$ nanocomposites. All the chemicals were used as obtained.

The synthesis process was initiated by mixing and stirring definite amounts of TEOS, in the double solvent of ethyl alcohol and distilled water in 10:3 (v/v). The pH of the solvent was controlled by $NH_3$. With careful monitoring of pH, different amounts of sonicated GNPs were added to the solution. The reaction was completed in 2 hours. The solution was dried at 373.15 K for 12 hours in an electric oven with post annealing session at 923.15K for 1 hour in a tube furnace. The $SiO_2$ was prepared under similar conditions without addition of GNPs. Three nanocomposites labelled as $S_x$, $S_y$, $S_z$ were prepared with different feed ratios of GNPs i.e., x = 40 mg, y = 80 mg, and z = 100 mg.

## B. PHOTOCATALYTIC DEGRADATION EXPERIMENT

A textile dye, MO was selected as a model pollutant to check the photocatalytic performance of graphene/$SiO_2$ nanocomposites. 0.03g of each photocatalyst was added to 15 $\mu$M aqueous MO solution. The pH of solution was adjusted to 3 using 1M $HNO_3$ (1M $NH_3$). 100 ml of each solution was subjected to the experiment. After establishing the adsorption-desorption equilibrium between photocatalyst and MO. Thereafter, the zero-time reading was recorded and the solution was exposed to UV light source (Type C, 90 Watt, emission peak 254 nm). 4 ml of each sample was withdrawn at regular intervals from all the solutions. The samples of the experimented solutions were analysed for absorbance immediately after centrifugation. Each experiment was repeated three times to ensure the accuracy of results.

## C. ANTIBACTERIAL ACTIVITY EXPERIMENT

To study the antibacterial performance, the 10 mg/ml of test samples ($SiO_2$, $S_x$, $S_y$, $S_z$) in sterile water were sonicated. The 200 $\mu$L of each sonicated solution was added to 5 ml Luria-Bertani Broth (LB) medium. This growth medium was added to 100 $\mu$L of the inoculum (bacterial culture in LB). The inoculated media containing the test sample was incubated at 37ºC for 24 h and all the bacteria were quantified by $OD_{600nm}$ at different intervals from 2 to 24 h.

### D. ANTICANCER TEST EXPERIMENT

Three model cell lines were selected to check the anticancer and cytotoxicity profile of prepared samples. In this regard two cancer cell lines, i.e., human cervical cancer (HeLa) cells derivative (HEp-2), human breast adenocarcinoma cell line (MCF-7), and a normal cell line i.e., human corneal epithelial cells (HCEC) were subjected to the experiment. All the chemicals used in cell toxicity experiments were purchased from Sigma Aldrich (USA).

To test the effect of prepared samples, all the cell lines were grown and maintained within incubator at 37ºC, 5% $CO_2$ in PRIM-1640 (Invitrogen-USA) supplemented with 5% antibiotics-antimycotic solution (GPPS) and 10% fetal bovine serum (FBS). In present work, the dosage dependent activity of prepared samples was checked against all cell lines. These cell lines were exposed to different doses i.e., 100 $\mu$g/ml to 400 $\mu$g/ml of each of the sample. The cells ($1 \times 10^5$) were seeded in 96-well culture plates and were grown in 5% $CO_2$ and humid atmosphere, for 48 hours at 37ºC in incubator both in the presence and absence (control) of test samples. The percent cell viability was obtained using the MTT assay.

### E. CHARACTERIZATIONS

The synthesized nanomaterials were characterized for their physical and chemical properties. The X-ray diffractograms were obtained by PANalytical X'Pert PRO diffractometer using Cu $K_\alpha$ radiation. A Shimadzu (IR Tracer-100) spectrometer was used for FTIR spectra. The morphology of prepared nanomaterials was studied by FE-SEM (MIRA3 TESCAN) and HRTEM (JOEL JEM 4000EX). Room temperature Raman spectra and photoluminescence measurements were taken by Ramboss using excitation laser of wavelength 514 nm and 325 nm wavelength, respectively. A Tristar 3020 Micromeritics (USA) Porosimetry analyser was used to measure surface area of samples using Braunauer-Emmet-Teller (BET) method. To study the photocatalytic properties, the UV-vis spectra were recorded by Perkin-Elmer (Lambda 25 UV).

## III. RESULTS AND DISCUSSIONS

### A. STRUCTURAL AND MORPHOLOGICAL INVESTIGATION

The crystallinity of prepared nanocomposites is analysed in the range of $20^o - 80^o$ as depicted in Figure 1. The presence of diffraction peak at $26^o$ corresponds to C (002) of the graphitic host matrix, and agrees with JCPDS-NO: 75-1621. The X-ray diffractograms show a broad halo in the region $2\theta = 20^o$–$30^o$. This obtuse peak indicates the amorphous nature of $SiO_2$ NPs. The X-ray diffractograms agree with previous study.[48]

Morphological investigations have been carried out to understand the shape and attachment between the constituent species. The Figure 2(a) illustrates that $SiO_2$ sample shows spherical shape, monodispersed particles with the diameter ranging from 230 nm - 260 nm. Figure 2 (b) and 2(c) reveal the microstructure of graphene/$SiO_2$ nanocomposites. The semi-transparent graphene sheets with anchored $SiO_2$ NPs are clearly visible in Figures 2(b) and (c), thereby confirming the intercalation of GNPs into few layers of graphene sheets. The insertion of $SiO_2$ NPs has thus served to reduce the van der Waal's interaction between the stacked GNPs. It is interesting to note that the particle size of $SiO_2$ has been greatly reduced during nanocomposite formation, which is logical and expected as well. Indeed, the confinement effect of graphene sheets has been observed in previous reports on graphene/metal oxide's nanocomposites.[49] Moreover, the large surface area provided by graphene during growth process serves to reduce the excessive aggregation of primary nuclei. An Energy Dispersive X-ray spectroscopy (EDX) coupled with FE-SEM was used to investigate the composition of pristine $SiO_2$ and graphene/$SiO_2$ nanocomposite. The presence of carbon, silicon and oxygen is confirmed by peaks [see insets in left bottom of Figure 2(a) and 2(b)].

To further investigate the nature of $SiO_2$ and nanocomposites, TEM and selected area electron diffraction (SAED) patterns are presented in the Figure 2(d), which demonstrates that the $SiO_2$ NPs are attached to graphene sheets. The SAED pattern supports the observation from X-ray diffractograms. The inset illustrates that the $SiO_2$ NPs attached to graphene sheets show amorphous nature. Their amorphous nature is retained after the formation of nanocomposites.

### B. GROWTH MECHANISM

The growth mechanism is explained schematically in the Figure 3. The graphene sheets offer the active sites and large surface area for the growth of $SiO_2$ NPs. This gives the advantage that a large surface area is available for the nucleation of the primary particles, thereby reducing the particle size of $SiO_2$ NPs as compared to pristine $SiO_2$. The confining effect of graphene sheets also contributes to stop nucleation process after a certain limit thus leading to the reduced particle size as compared to pristine $SiO_2$.[48,49] The $SiO_2$ spheres have served to intercalate the graphene sheets as they get drafted on graphene and reduce the van der Waal's interaction between the sheets.

### C. FTIR AND RAMAN ANALYSIS

The FTIR spectroscopic curves are shown in Figure 4. A band at 460 $cm^{-1}$ can be assigned to the Si-O-Si bending vibrations. The band at 812 $cm^{-1}$ originates due to Si-O symmetric bending vibrations, where –O vibrations are perpendicular to the Si-Si bond line. However, the band in the range 1045-1107 $cm^{-1}$ is due to parallel vibrations of oxygen atom in either direction in Si-O-Si linkage. Thus, it manifests asymmetric mode of Si-O-Si bond. The adsorbed water molecules manifest themselves by a band around 1615 $cm^{-1}$. The chemical bond Si-OH appears as a band around 957cm$^{-1}$ and 3458 cm$^{-1}$. The graphene/SiO$_2$ nanocomposites show significant difference with the presence of a broad band in the range 1014-1303 cm$^{-1}$. The C-Si bond manifests itself in this region around 1260 cm$^{-1}$. Thus, it can be concluded that SiO$_2$ fabrication modified the surface of graphene. The FTIR results accord well with previous studies.[43, 48]

Raman spectroscopy is the basic characterization tool to identify graphene and its nanocomposites. The Raman spectra obtained from 1000 cm$^{-1}$ to 2000 cm$^{-1}$ are presented in Figure 5. All samples show D band, associated with defects, located around 1350 cm$^{-1}$. The defect band arises due to termination of sheet at edges and attachment of particles on graphene. The E$_{2g}$ mode arises due to first order scattering and is manifested as G band, located at 1587 cm$^{-1}$ in case of GNPs. These observations agree with previous reports on graphene's Raman spectra.[50,51] The nanocomposites S$_x$, S$_y$, S$_z$ have G-band located at 1602 cm$^{-1}$, 1599 cm$^{-1}$, and 1602 cm$^{-1}$. An overall shift towards higher wavenumber of G-band is observed in all the nanocomposites as compared to GNPs. This shift is due to the charge transfer between GNPs and SiO$_2$ NPs, and is an indicator of electrostatic interaction between the two-constituent species. This large shift illustrates the strong attraction between the two constituent phases of nanocomposites. The Raman results confirm the formation of GNPs and SiO$_2$ nanocomposites.[52]

## D. PHOTOLUMINESCENCE AND SURFACE AREA ANALYSIS

To monitor the optical changes due to structural defects, a comparison of photoluminescence spectra of SiO$_2$ and graphene/SiO$_2$ nanocomposites is presented in Figure 6. The incomplete Si-O-Si tetrahedral network formation on the surface of graphene may lead to several structural defects. The SiO$_2$ nanoparticles show emission peaks in the visible light region. A weak band near UV region~355.6 nm (3.48 eV) is contributed by silanol groups (-OH related groups). Green emission~ 512nm (2.43 eV) is observed. A very prominent band~ 409 nm (3.04eV) is observed related to violet emission.[53,54]

The comparison of the PL results of graphene/$SiO_2$ nanocomposites and $SiO_2$ indicates that the intensity of all emission peaks quenches significantly. This suppression is attributed to the presence of graphene, which acts as an acceptor of electrons in the nanocomposite.[55] Graphene sheets provide an additional path for the conduction electrons of $SiO_2$. The suppression of PL intensity indicates the decrease in carriers' recombination. This quenching behaviour agrees with the previous reports as well suggests the potential photocatalytic use of prepared nanocomposites.

The surface area is examined by $N_2$ adsorption-desorption isotherms. The observed values are tabulated in Table 1. The $SiO_2$ particles with relatively larger particle size, possess the smallest BET surface area. With the increment of graphene content, the surface area has been increased significantly. Since, theoretically, graphene possesses a surface area $\approx 2600$ $m^2g^{-1}$. The highest BET surface area is observed for $S_z$ i.e., 146.52 $m^2g^{-1}$.

| Sample | Surface area ($m^2 g^{-2}$) | Pore volume ($cm^3 g^{-1}$) | Average pore size ($\mathring{A}$) |
|---|---|---|---|
| $SiO_2$ | 6.3547 | 0.019405 | 55.157 |
| $S_x$ | 16.0839 | 0.0210532 | 46.234 |
| $S_y$ | 30.2417 | 0.040561 | 35.904 |
| $S_z$ | 146.5199 | 0.393357 | 33.021 |

**Table 01**

## E.  PHOTOCATALYTIC DEGRADATION OF MO

Finally, to evaluate the performance of graphene/SiO$_2$ nanocomposites, MO was employed as water contaminant in photo induced dye-degradation experiments.

To study the impact of increased graphene concentration on MO degradation, four experiments were conducted using catalysts SiO$_2$, S$_x$, S$_y$, and S$_z$. Figures 7(a)-(d) show the absorption spectra detailing degradation of MO. The adsorption of catalysts (in the dark) on degrades the dye molecules slightly. The photodegradation efficiencies of MO are presented in Figure 8 using different photocatalysts. In the presence of SiO$_2$, 46% photodegradation is achieved in 160 min. The catalyst with minimal graphene content, i.e., S$_x$ achieves 87.2% efficiency in 160 min. The catalyst S$_y$ has shown 92% photodegradation of MO in 160 min. The photocatalytic efficiency of graphene/SiO$_2$ is maximized at the optimal graphene content in the photocatalyst S$_z$ which shows 99% photodegradation in considerably reduced time. The spectrum becomes flat only in 100 min. The photocatalytic efficiency obtained in present case is much better than previous report on photocatalytic activity of SiO$_2$ NPs with Au/Ag doping.[21, 22] The mechanism of photocatalytic activity is explained below.

The photons of UV light falling on the SiO$_2$ nanoparticles, excite its valence band electrons to the conduction band and produce e$^-_{cb}$ − h$^+_{vb}$ pairs. The number of e$^-_{cb}$ − h$^+_{vb}$ pairs increases gradually with time. The graphene attached to SiO$_2$ NPs, being a good acceptor of electrons, provides trapping sites for e$^-_{cb}$. This delays the recombination of e$^-_{cb}$ − h$^+_{vb}$ pairs. Meanwhile the e$^-_{cb}$ may also interact with the dissolved O$_2$ to produce O$^-_2$ species, which may further produce several reactive oxygen species (ROS) as mentioned in the following equations. The holes in the valence band of SiO$_2$ contribute towards generation of $^\bullet$OH radicals.[21,22] These species attack the ring of azo dye, MO, by completely opening its ring structure. This ultimately results into the mineralization of dye.[37] The mechanism is schematically illustrated in Figure 9.

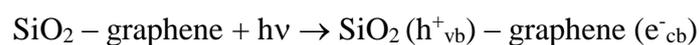

$$SiO_2 - graphene + h\nu \rightarrow SiO_2 (h^+_{vb}) - graphene (e^-_{cb})$$

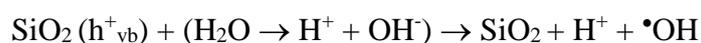

$$SiO_2 (h^+_{vb}) + (H_2O \rightarrow H^+ + OH^-) \rightarrow SiO_2 + H^+ + {}^\bullet OH$$

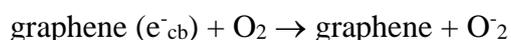

$$graphene (e^-_{cb}) + O_2 \rightarrow graphene + O^-_2$$

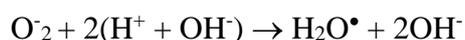

$$O^-_2 + 2(H^+ + OH^-) \rightarrow H_2O^\bullet + 2OH^-$$

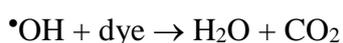

$${}^\bullet OH + dye \rightarrow H_2O + CO_2$$

In the case of bare $SiO_2$ particles, the $e^-_{cb} - h^+_{vb}$ pairs formed on the surface of $SiO_2$ recombine quickly. Only very few carriers can be trapped on surface states of $SiO_2$ particles, which may further initiate the dye degradation. The amount of ROS generated in the process strictly controls the dye degradation. Fewer ROS can react with adsorbed dye molecules. In this case absence of graphene leads to quick recombination of $e^-_{cb} - h^+_{vb}$ pairs. That is why $SiO_2$ particles have shown low photocatalytic activity as compared to graphene/$SiO_2$ nanocomposites.

Notably, all the nanocomposites have not shown the similar photoactivity for the degradation of MO. It indicates that the importance of optimum addition ratio of graphene in nanocomposites. The explanation for the graphene content dependence of photocatalytic performance of the nanocomposites is elucidated below.

Incorporating graphene in nanocomposites seems to promote electron trapping. This is due to exceptional conductivity of graphene. The retardation in recombination of UV light generated charge carriers in $SiO_2$ by introducing graphene nanoplatelets gets support from photoluminescence spectra (Figure 6). PL findings suggest that the quenching of intensity is due to inhibition of charge carrier's recombination in nanocomposites. As electrons are accepted by graphene lying adjacent to the $SiO_2$ NPs. This inhibition ultimately consequences the efficient photocatalytic performance of nanocomposites as compared to $SiO_2$ alone. The highest quenching is observed in $S_z$ which supports the photocatalytic results. The nanocomposite having highest quenching has shown most efficient photocatalytic activity by 99% degradation of MO in 100 minutes. The surface area of sample is likely to affect the performance of photocatalyst. The high surface area of graphene/$SiO_2$ nanocomposites (Table 1) offers more area (as compared to $SiO_2$) for better adsorption of MO molecules. Therefore, a better contact between the dye and photocatalyst is developed. Hence, the synergistic effect created by conducting graphene and high surface area is helpful towards fast and efficient degradation of dye.

The reaction kinetics of experiment give a better insight to the photocatalytic activity. The reactions kinetics for photocatalysis can be described on the bases of Langmuir-Hinshelwood model[56] as presented in Figure 10. The UV light induced degradation of MO can be well ascribed by pseudo-first order kinetics. The rate equation that describes the reaction is $\ln(C_t/C_o) = kt$, where $C_t$ and $C_o$ are the dye concentration at time t and the initial concentration respectively, and k = apparent rate constant. The apparent rate constant determined for different catalysts are 0.003 min$^{-1}$ ($SiO_2$), 0.014 min$^{-1}$($S_x$), 0.015 min$^{-1}$ ($S_y$), and 0.04 min$^{-1}$($S_z$). These results show that the rate constant has been increased significantly by increasing graphene content. The graphene/$SiO_2$ nanocomposite with maximum graphene loading has the highest apparent rate constant (an order of magnitude higher than that of bare $SiO_2$ particles) and therefore it exhibits excellent photocatalytic activity. Hence, it can be concluded that $S_z$ nanocomposite is photocatalytically most active than pristine $SiO_2$ and graphene/$SiO_2$ nanocomposites with low graphene content for the degradation of MO. This observation is ascribed to the synergistic effect of excellent electron acceptor nature of graphene and higher surface area of graphene/$SiO_2$ nanocomposites.[10,23]

The recyclability tests are very important for the practical use of photocatalyst. The recycling performance of $S_z$ was evaluated for consecutive three cycles. It is illustrated by inset in Figure 10 that there is negligible loss in photocatalytic activity of $S_z$. The $S_z$ nanocomposite shows excellent performance even after three continuous cycles of activity. Therefore, it may be recommended as efficient alternative of traditional photocatalysts.

### F. ANTIBACTERIAL STUDY

To investigate the effect of graphene/$SiO_2$ nanocomposites on Gram negative and Gram positive bacteria, the time kill assay was conducted for three model bacterial strains i.e., *Escherichia coli (E. coli),* Methicillin resistant *Staphylococcus aureus (S. aureus),* and *Pseudomonas aeruginosa (P. aeruginosa)* in the absence and presence all samples. The growth inhibition rate was determined by observing the optical density (600 nm) at different intervals in the total incubation time (24 h). The results obtained from the experiment are presented by the bacterial growth inhibition curves in Figures 11(a)-(c). The control sample in Figure 11(a)-(c) presents the untreated bacterial strains under observation for comparison purposes. The experimental findings suggest the inhibition of bacterial growth to a significant extent. It is observed that bacterial growth inhibition is a strong function of graphene loading in the nanocomposites. The growth of *S. aureus* has been inhibited up to 26.32%, 30.17%, 50.30%, and 75.40% by $SiO_2$, $S_x$, $S_y$, and $S_z$ respectively. For *E. coli,* the growth inhibition rates are 17.00%, 30.00%, 32.23%, 51.80% for $SiO_2$, $S_x$, $S_y$, and $S_z$ respectively. The growth of *P. aeruginosa* has been inhibited upto 17.75%, 40.52%, 40.77%, and 48.97% by $SiO_2$, $S_x$, $S_y$, and $S_z$ respectively. The sample with no graphene contents i.e., $SiO_2$ shows the minimum inhibition of bacterial growth for all the bacterial strains. The increase of graphene loading in nanocomposites decreases the number of viable cells. The maximum growth inhibition rate is achieved in nanocomposite with maximum graphene content. The sample with maximum graphene loading ($S_z$) is found to possess excellent antibacterial properties for the growth inhibition of *S. aureus*. Graphene/SiO$_2$ nanocomposites have stopped around 50% growth of Gram negative bacterial strains. These results are much better than the previous reports, where 47% and 49.5 ± 4.8% growth inhibition of *E. coli*, and 34% growth inhibition of *S. aureus* was achieved by using rGO and graphene films.[57-59]

Several research reports are available, documenting the possible mechanisms for growth inhibition of bacterial strains by carbon nano tubes, fullerenes, and graphene family nanostructures.[60-65] But the exact mechanism explaining the loss of bacterial integrity is still a researchable topic. One of the suggested mechanism is the destruction of bacterial membrane induced by direct contact between sheet like structure of graphene based materials and bacteria. This mechanism has been proposed previously for GO, rGO, and CNTs.[58] In present case, intermingling of bacteria and planar graphene may be thought to induce irreversible destruction of bacterial membrane. The planes and sharp edges of graphene nanosheets produce significant stress on cell membranes. These edges serve as cutters for rupturing the bacterial cell membranes which induce cell death by leakage of cytoplasmic content. Additionally, at the same time the normal respiratory functioning of bacteria is strongly dependent on electronic charge transport between the cell and mitochondrial membranes in respiratory chain reactions. The physical contact between bacteria and graphene/SiO$_2$ nanocomposites may result in Schottky barrier formation, as previously reported for graphene on SiO$_2$ substrate.[66] As graphene is an excellent electron acceptor so it can be speculated that cell membranes may lose their electrons, which are eventually transported to graphene. In this manner, a charge imbalance is created in bacterial cells which leads to cell death.[66,67]

However, the differential toxicity of samples towards Gram positive and Gram negative bacteria can be explained on the bases of their outer membrane's composition. The Gram-negative bacteria have a complex double membrane's structure which is less penetrable as compared to single membrane of Gram positive bacteria. This difference, mainly arising due to chemical composition of membranes of the two classes of bacteria, develops their differential resistance.[68,69] Hence, this feature makes *S. aureus* an easy target for graphene/SiO$_2$ nanocomposites.

### G. CYTOTOXICITY ANALYSIS

It is highly desirable to evaluate the cytotoxic response of a nanoplatform against various cell lines before its recommendation for drug attachment to kill the cancer cells. The dosage dependent toxicity of prepared samples towards different cell lines (HCEC, MCF-7 and HEp-2) was evaluated. The absorbance at 550 nm was recorded and % cell viability was obtained. The untreated cell lines were considered as control. A comparison between control and treated cell lines show that the number of viable cells doesn't decrease after the exposure to prepared samples. The dosage increment from100 to 400 $\mu$g/ml also doesn't create the toxic effects both on cancer cell lines and normal cell lines. No apoptosis or cell death is induced using any of the prepared samples at exposure concentrations as high as 400 μg/ml and exposure time as much long as 48 hours. These findings establish that SiO$_2$ particles and graphene/SiO$_2$ nanocomposites are non-toxic for used cell lines. Their non-toxicity towards healthy cells is an excellent indication for their utilization as a nanoplatform for efficient drug delivery. The prepared graphene/SiO$_2$ nanocomposites can be functionalized with anti-cancer drugs like doxorubicin as previously done for carbon nanoparticles. The biofunctionalization of graphene with SiO$_2$ and doxorubicin is expected to target the cancerous cell lines by making use of combined mechanism of photothermal therapy and chemotherapy.[42] Moreover, the slight modification in morphology of prepared SiO$_2$ (that is making them mesoporous) and attachment of doxorubicin can target glioma.[70] This nanoplatform can be modified with hypocrellin A suggesting their efficacy in drug delivery for photo dynamic cancer treatment and bio-imaging.[71] We conclude that the cytotoxicity analysis presented in this study, opens the doors towards covalent attachment of anti-cancerous drugs on prepared graphene/SiO$_2$ nanocomposites.

## IV.    CONCLUSIONS

High surface area graphene/SiO$_2$ nanocomposites are synthesized successfully by a simple chemical route. The composite developed with retention of the exceptional intrinsic properties of graphene leads to achieve the outstanding photocatalytic performance with 99% degradation of MO under UV light illumination. Graphene loaded with SiO$_2$ shows excellent antibacterial activities i.e., 75% growth inhibition of *S. aureus* and ≈ 50% loss of *E. coli*, and *P. aeruginosa*. The prepared nanocomposites show no toxicity towards normal cells (HCEC), human cervical cancer cells (HEp-2), and breast cancer cells (MCF-7) which suggests their possible use as nano drug carriers to target cancerous sites. Our experimental findings greatly recognize graphene/SiO$_2$ nanocomposites for their utilization towards waste water treatment and biomedical applications.

**ACKNOWLEDGEMENTS**

This work was funded by the Higher Education Commission of Pakistan (HEC) NRPU (Grant No: 20-4861/R & D/ HEC/14) to Dr. Javed Iqbal and Higher Education Commission of Pakistan (HEC) IRSIP (Grant No: 1-8/HEC/HRD/2016/5995 PIN: IRSIP 32 PSc 04) to Aqsa Arshad. The authors are thankful to Dr. Ian Terry and Dr. Mahavir Sharma, Department of Physics, Durham University, UK, for the useful discussions. The plagiarism check of the article (using Turnitin ID: 801482706) as recommended by HEC Pakistan, shows 2% similarity index which is well below the permissible range.

**REFERENCES**

[1]    A. Hameed, Z. Shafiq, M. Yaqub, M. Hussain, M. A. Hussain, M. Afzal, M. N. Tahir and M. M. Naseer, New J. Chem. **39,** 9351-9357 (2015).

[2]    L. Zhao, H. Wang, K. Huo, L. Cui, W. Zhang, H. Ni, Y. Zhang, Z. Wu, P. K. Chu, Biomaterials **32**, 5706–5716 (2011).

3     M. Zhang, K. Zhang, B. D. Gusseme, W. Verstraete, Water Res. **46**, 2077–2087 (2012).

4     P. Prombutara, Y. Kulwatthanasal, N. Supaka, I. Sramala, S. Chareonpornwattana, Food Control **24**, 184–19 (2012).

5     L. Cheng, M. D. Weir, H. H. K. Xu, J. M. Antonucci, A. M. Kraigsley, N. J. Lin, S. Lin-Gibson, X. Zhou, Dent. Mater. **28**, 561–572 (2012).

6     F. Anam, A. Abbas, K. Mun Lo, Z. Rehman, S. Hameed and M. M. Naseer, New J. Chem. **38**, 5617-5625 (2014).

7     Kenry, P. K. Chaudhuri, K. P. Loh, C. T. Lim, ACS Nano **10**, 3424–3434 (2016).

8     H. R. Jafry, M. V. Liga, Q. Li, A. R. Barron, Environ. Sci. Technol. **45**, 1563–1568 (2011).

9     B. Czech, W. Buda, Environ. Res.**137**, 176–184 (2015).

10    T. Kavitha, A. I. Gopalan, K. P. Lee, S. Y. Park, Carbon **50**, 2994–3000 (2012).

11    Y. Li, W. Zhang, J. Niu, Y. Chen, ACS Nano **6**, 5164–5173 (2012).

12    G. A. Seisenbaeva, M. P. Moloney, R. Tekoriute, A. H. Dessources, J. M. Nedelec, Y. K. Gun'ko, V. G. Kessler, Langmuir **26**, 9809–9817 (2010).

13    A. Popat, S. B. Hartono, F. Stahr, J. Liu, S. Z. Qiao, G. Q. M. Lu, Nanoscale **3**, 2801−2818 (2011).

14    X. Yang, Z. Shen, B. Zhang, J. Yang, W. X. Hong, Z. Zhuang, J. Liu, Chemosphere **90**, 653−659 (2013).

15    R. D. C. Solatani, A. R. Khataee, M. Safari, S.W. Joo, Int. Biodeterior. Biodegrad. **85**, 383−391 (2013).

16    X. Li, J. Zhang, H. Gu, Langmuir **10**, 6099−6106 (2011).

17    B. Bharti, J. Meissner, G. H. Findenegg, Langmuir **27**, 9823−9833 (2011).

18    Y. Badr, M. G. A. E. Wahed, M. A. Mahmoud, J. Haz. Mat. **15**, 245–253 (2008).

19    Y. Badr, M. A. Mahmoud, J. Phys. Chem. Solids **68**, 413–419 (2007).

20    L. F. D. Oliveira, K. Bouchmella, K. D. A. Gonçalves, J. Bettini, J. Kobarg, M. B. Cardoso, Langmuir **32**, 3217–3225 (2016).

21    B. Xu, Y. Ju, Y. Cui, G. Song, Y. Iwase, A. Hosoi, Y. Morita, Langmuir **30**, 7789–7797 (2014).

22    C. Giménez, C. D. Torre, M. Gorbe, E. Aznar, F. Sancenón, J. R. Murguía, R. Martínez-Máñez, M. D. Marcos, P. Amorós, Langmuir **31**, 3753–3762 (2015).

23    X. Li, S. Yang, J. Sun, P. He, X. Xu, G. Ding, Carbon **78**, 38–48 (2014).

24    G. Jiang, Z. Lin, C. Chen, L. Zhu, Q. Chang, N. Wang, W. Weia, H. Tang, Carbon **49**, 2693–2701 (2011).

25    V. Kumara, N. Bahadur, D. Sachdeva, S. Guptaa, G. B. Reddyb, R. Pasricha, Carbon **80**, 290-2304 (2014).

26    N. O. Ramoraswi, P. G. Ndungu, Nanoscale Res. Lett. **10**, 427-443 (2015).

27    K. S. Novoselov, A. K. Geim, S. V. Morozov, D. Jiang, Y. Zhang, S. V. Dubonos, I. V. Grigorieva1, A. A. Firsov, Science **306**, 666-669 (2004).

28    N. A. H. Castro, F. Guinea, N. M. R. Peres, K. S. Novoselov, A. K. Geim, Rev. Mod. Phys. **81**, 109–162 (2009).

29    C. N. R. Rao, A. K. Sood, K. S. Subrahmanyam, A. Govindaraj, Angew. Chem. Int. Ed. **48**, 7752–7777 (2009).

30    A. K. Geim, K. S. Novoselov, Nat. Mater. **6**, 183–191 (2007).

31    R. Muszynski, B. Seger, P. V. Kamat, J. Phys. Chem. C. **112**, 5263–5266 (2008).

32    Y. Y. Wen, H. M. Ding, Y. K. Shan, Nanoscale **3**, 4411–4417 (2011).

33    H. Seema, K. C. Kemp, V. Chandra, K. S. Kim, Nanotechnol. **23**, 355705–355712 (2012).

34  S. Stankovich, D. A. Dikin, G. H. B. Dommett, K. M. Kohlhaas, E. J. Zimney, E. A. Stach, R. D. Piner, S. T. Nguyen, R. S. Ruoff, Nature **442**, 282–286 (2006).

35  T. S. Sreeprasad, S. M. Maliyekkal, K. P. Lisha, T. Pradeep, J. Hazard. Mat. **186**, 921–31 (2011).

36  C. Xu, X. Wang, J. Zhu, J. Phys. Chem. C **112**, 19841–19845 (2008).

37  A. Arshad, J. Iqbal, M. Siddiq, Q. Mansoor, M. Ismail, F. Mehmood, M. Ajmal, and Z. Abid, J. Appl. Phys. **121**, 024901 (2017).

38  Y. Gao, X. Pu, D. Zhang, G. Ding, X. Shao, J. Ma, Carbon **50**, 4093–4101 (2012).

39  [a] S. C. Smith, D. F. Rodrigues, Carbon **91**, 22-143 (2015).
    [b] X. Liang, S. Liu, X. Song, Y. Zhu, S. Jiang, Analyst **137**, 5237–5244 (2012).

40  H. M. Hegab, A. E. Mekawy, L. Zou, D. Mulcahya, C. P. Saint, M. G. Markovic, Carbon **105**, 362-376 (2016).

41  L. Yang, L. Wang, M. Xing, J. Lei, J. Zhang, Appl. Catal. B: Environ. **180**, 106–112 (2016).

42  X. Tu, L. Wang, Y. Cao, Y. Ma, H. Shen, M. Zhang, Z. Zhang, Carbon **97**, 35–44 (2016).

43  C. W. Lee, K. C. Roh, K. B. Kim, Nanoscale **5**, 9604-9608 (2013).

44  Z. M. Wang, W. Wang, N. Coombs, N. Soheilnia, G. A. Ozin, ACS Nano **4**, 7437–7450 (2010).

45  S. Watcharotone, D. Z. Dikin, S. Stankovich, R. Piner, I. Jung, G. H. B. Dommett, G. Evmenenko, S. E. Wu, S. F. Chen, C. P. Liu, S. T. Nguyen, R. S. Ruoff, Nano Lett. **7**, 1888-1892 (2007).

46  D. Yin, B. Liu, L. Zhang, M. Wu, J. Biomed. Nanotechnol. **8**, 458-464 (2012).

47  L. Kou, C. Gao, Nanoscale **3**, 519–528 (2011).

48  N. D. Singho, M. R. Johan, Int. J. Electrochem. Sci. **7**, 5604–5615 (2012).

49    B. Zhao, J. Song, P. Liu P, W. Xu, T. Fang, Z. Jiao, H. Zhang, Y. Jaing, J. Mat. Chem. **21**, 18792-18798 (2011).

50    T. Tsukamoto, K. Yamazaki, H. Komurasaki, T. Ogino, J. Phys. Chem. C **116**, 4732-4737 (2012).

51    A. C. Ferrari, J. C. Meyer, V. Scardasi, M. Lazzeri, F. Mauri, S. Piscanec, D. Jiang, K. S. Novoselov, S. Roth, A. K. Geim, Phys. Rev. Lett. **97**, 187401-187404 (2006).

52    A. K. Rai, J. Gim, L. T. Anh, J. Kim, Partially reduced $Co_3O_4$/graphene nanocomposite as an anode material for secondary lithium ion battery, Electrochimica Acta **100**, 63-71 (2013).

53    M. Jafarzadeh, I. A. Rahman, C. S. Sipaut, Ceramics Int. **36**, 333–338 (2010).

54    I. A. Rahman, P. Vejayakumaran, C. S. Sipaut, J. Ismail, C. K. Chee, Mat. Chem. Phy. **114,** 328–332 (2009).

55    X. An, J. C. Yu, Y. Wang, Y. Hu, X. Yu, G. Zhang, J. Mat. Chem. **22**, 8525–8531 (2012).

56    K. V. Kumar, K. Porkodi, F. Rocha, Cat. Comm. **9**, 82-84 (2008).

57    W. Hu, C. Peng, W. Luo, M. Lv, X. Li, D. Li, Q. Huang, C. Fan, ACS Nano **4**, 4317–4323 (2010).

58    S. Liu, T. H. Zeng, M. Hofmann, E. Burcombe, J. Wei, R. Jiang, J. Kong, Y. Chen, ACS Nano **5**, 6971–6980 (2011).

59    L. Dellieu, E. Lawarée, N. Reckinger, C. Didembourg, J. J. Letesson, M. Sarrazin, O. Deparis, J. Y. Matrouleb, J. F. Colomer, Carbon **84**, 310–316 (2015).

60    E. Nakamura, H. Isobe, Acc. Chem. Res. **36**, 807-815 (2003).

61    B. Sitharaman, L. J. Wilson, J. Biomed. Nanotech. **3**, 342-52 (2007).

62    L. Lacerda, A. Bianco, M. Prato, K. Kostarelos, Adv. Drug Deliv. Rev. **58**, 1460-1470 (2006).

[63]  Z. Liu, S. Tabakman, K. Welsher, H. Dai, Nano Res. **2**, 85-120 (2009).

[64]  O. Akhavan, E. Ghaderi, ACS Nano **4**, 5731–5736 (2010).

[65]  Y. B. Zhang, S. F. Ali, E. Dervishi, Y. Xu, Z. R. Li, D. Casciano, A. S. Biris, ACS Nano **4**, 3181–3186 (2010).

[66]  J. Li, G. Wang, W. Zhu, M. Zhang, X. Zheng, Z. Di, X. Li, X. Wang, Sci. Rep. **4**, 4359 (2014).

[67]  H. W. Harris, M. Y. El-Naggar, O. Bretschger, M. J. Ward, M. F. Romine, A. Obraztsova, K. H. Nealson, PNAS **107**, 326–331 (2010).

[68]  J. K. Krishnamoorthy, M. Veerapandian, L. H. Zhang, K. Yun, S. J. Kim, J. Phys. Chem. C **116**, 17280−17287 (2012).

[69]  A. M. Jastrzębska, P. Kurtycz, A. R. Olszyna, J. Nanopart. Res. **14**, 1320-41 (2012).

[70]  Y. Wang, K. Wang, J. Zhao, X. Liu, J. Bu, X. Yan, R. Huang, J. Am. Chem. Soc. **135**, 4799-4804 (2013).

[71]  L. Zhou, L. Zhou, X. Ge, J. Zhou, S. Wei, J. Shen, Chem. Comm. **51,** 421-424 (2015).

**Table Legends**

**Table 1**. Parameters obtained from $N_2$ adsorption-desorption isotherms.

**Figure Captions**

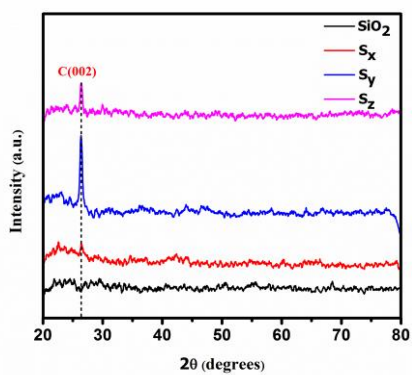

**Figure 1.** X-ray diffractograms depicting $SiO_2$ and graphene/$SiO_2$ nanocomposites.

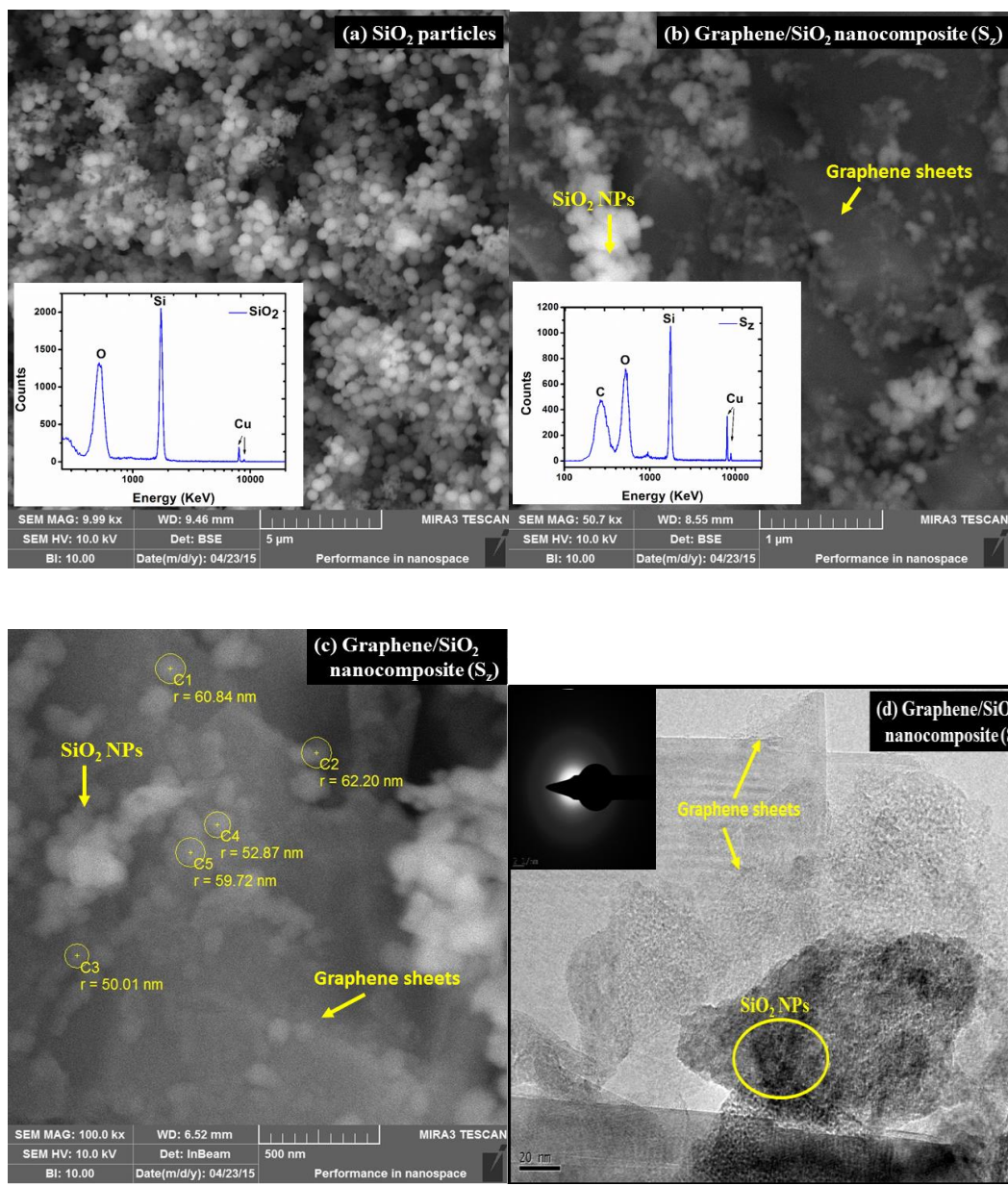

**Figure 2.** (a) SEM micrograph of monodispersed SiO₂ particles, inset is EDX spectrum of SiO₂, (b) and (c) graphene/SiO₂ nanocomposite, inset represents EDX spectrum, and (d) TEM image of graphene/SiO₂ nanocomposite (S$_z$), inset shows the SAED pattern of SiO₂ NPs attached to graphene sheet.

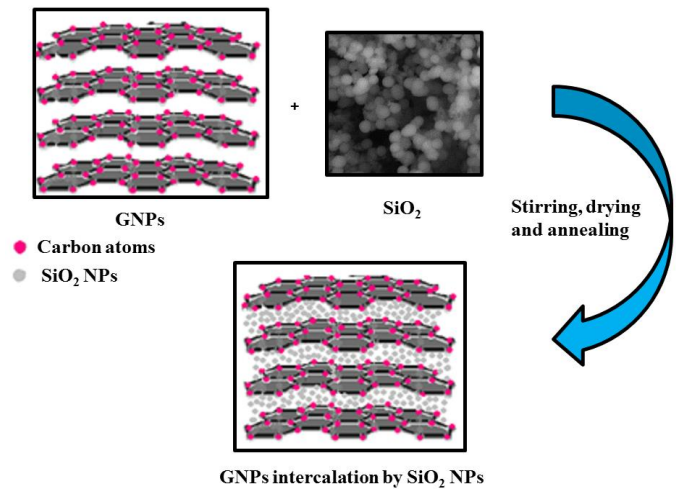

**Figure 3.** Schematic illustration of intercalation of graphene nanoplatelets by $SiO_2$ NPs.

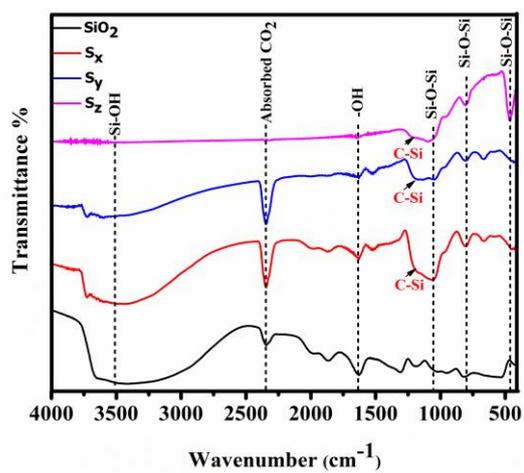

**Figure 4.** FTIR spectra of $SiO_2$ and graphene/$SiO_2$ nanocomposites.

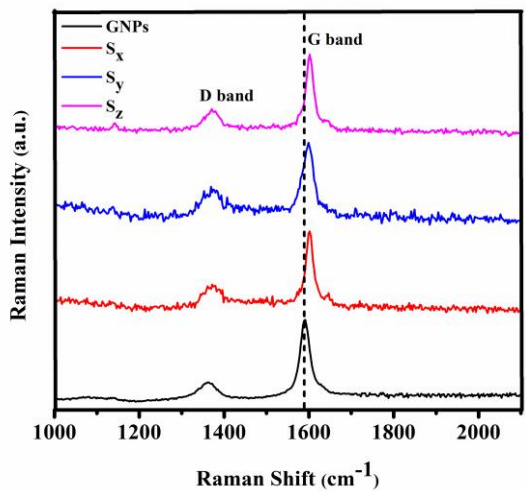

**Figure 5.** Raman spectra depicting the formation of graphene/SiO$_2$ nanocomposites.

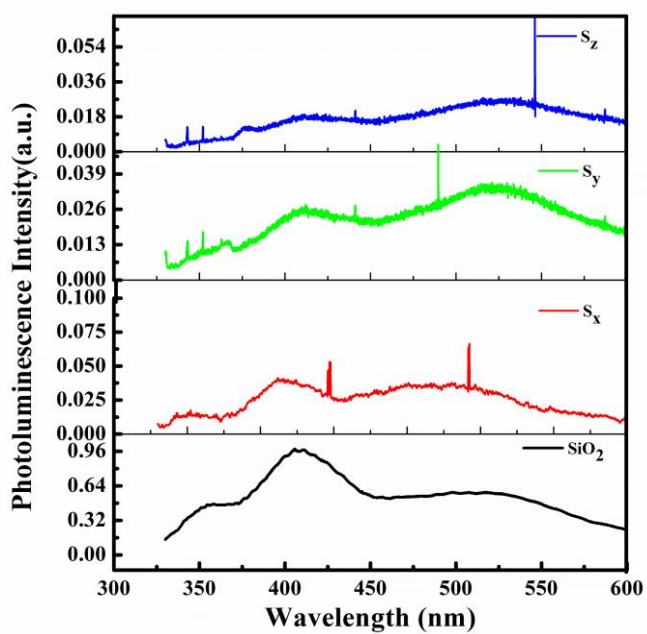

**Figure 6**. PL spectra of SiO$_2$ and graphene/SiO$_2$ nanocomposites.

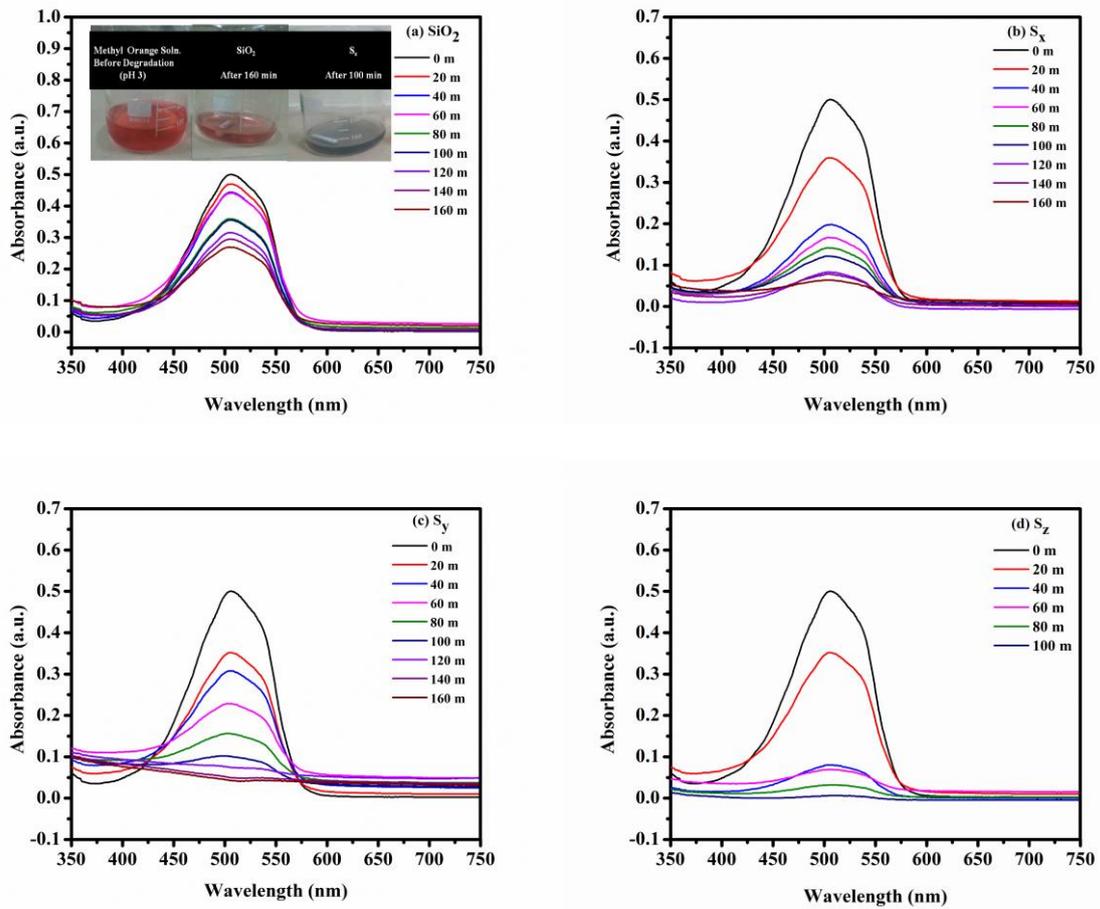

**Figure 7 (a-d)**. UV-Vis absorbance spectra of methyl orange in presence of different photocatalysts i.e., $SiO_2$, $S_x$, $S_y$, and $S_z$.

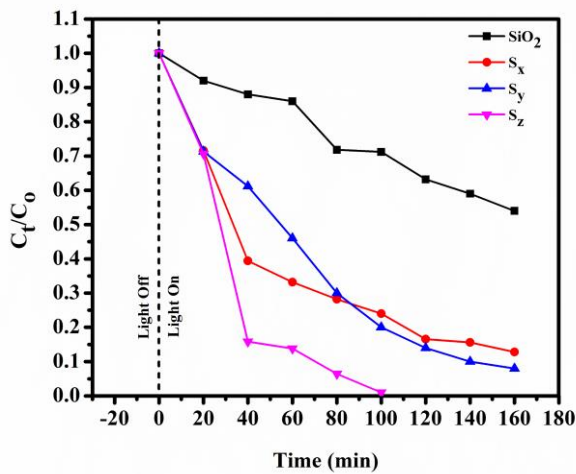

**Figure 8**. Photodegradation of methyl orange by $SiO_2$ and graphene/$SiO_2$ nanocomposites.

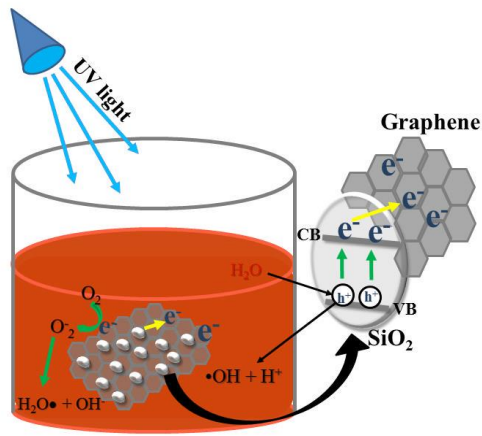

**Figure. 9**. Proposed mechanism for UV light induced catalysis of MO using graphene/SiO₂ nanocomposites.

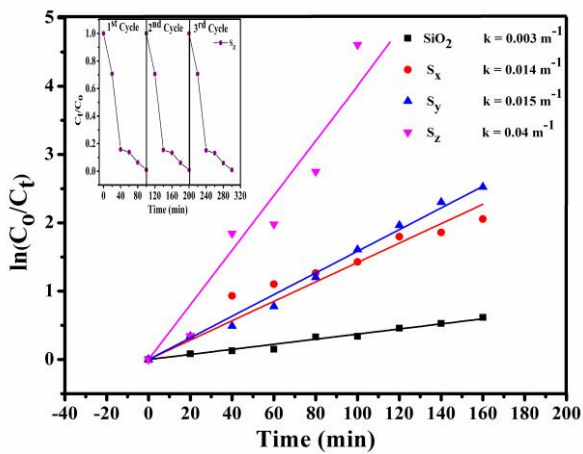

**Figure 10**. Pseudo first order rate kinetics for photocatalytic reactions by SiO₂ and graphene/SiO₂ nanocomposites. Inset: recyclability curves of graphene/SiO₂ nanocomposite (S$_z$) depicting its excellent performance in three consecutive cycles of photocatalysis.

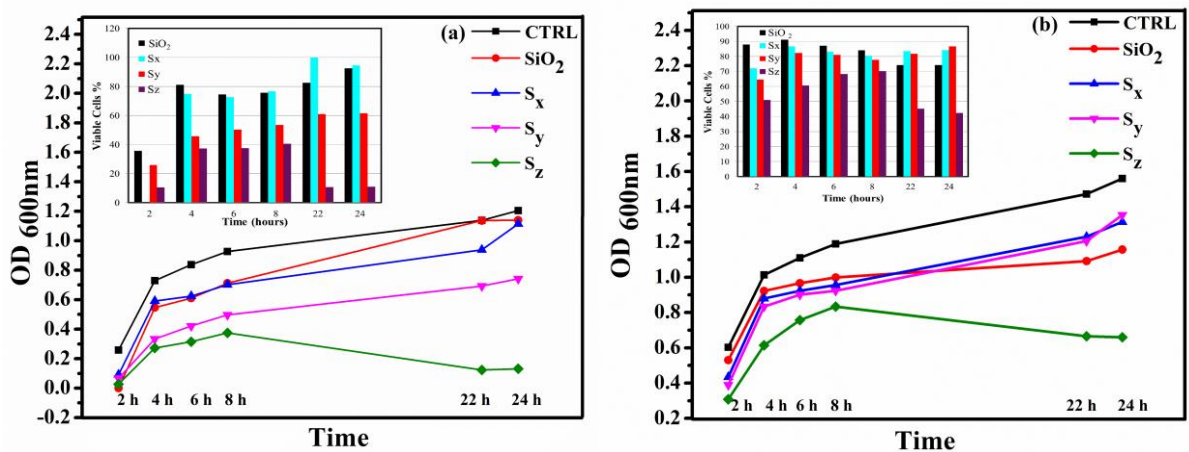

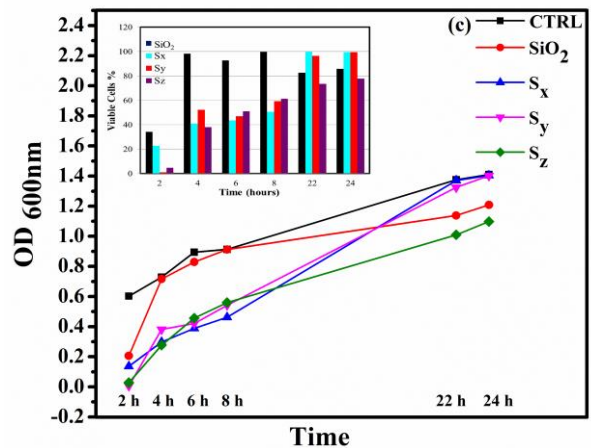

**Figure 11**. Bacterial growth inhibition curves obtained via MTT assay for (a) Methicillin resistant *S. aureus,* (b) *E. coli* and (c) *P. aeruginosa.* Insets: % cell viability of respective bacterial strains.